\documentclass[%
floatfix,
 aps,
 prc,%
 amsmath,amssymb,amstex,
  preprint,%
  showpacs,%
]{revtex4-1}
\usepackage{graphicx}
\usepackage{dcolumn}
\usepackage{bm}
\usepackage{soul}
\usepackage{color}
\usepackage{MnSymbol}
\usepackage{cancel}
\usepackage{threeparttable}
\usepackage{longtable}
\usepackage{multirow}
\usepackage{lineno}
\usepackage{listings}

\usepackage{paralist}
\usepackage{hyperref}
\definecolor{darkblue}{rgb}{0.0,0.0,0.66}
\definecolor{darkgreen}{rgb}{0.0,0.5,0.0}
\hypersetup{colorlinks,breaklinks,
            linkcolor=darkblue,urlcolor=darkblue,
            anchorcolor=darkblue,citecolor=darkblue}

\def \mev{\text{MeV}}

\def\gps{$\gamma$/s}

\def\ap{$\alpha$-particle}
\def\apa{$\alpha+\alpha$}
\def\aatobe{$\alpha + \alpha~\rightarrow~^{8}$Be}
\def\aantoben{$\alpha + \alpha + n~\to~^9$Be}
\def\sup#1{$^{#1}$}

\def\ben{$^{9}$Be}
\def\bee{$^{8}$Be}

\def\dtwoO{D$_{2}$O}

\def\aan{$\left<\alpha\alpha n\right>$}

\def\sv{$\left<\sigma v\right>$}
\def\svmm{\left<\sigma v\right>}

\def\forward{$^{9} \text{Be}  +  \gamma ~\rightarrow ~^{8} \text{Be}  +  n$}
\def\reverse{$^{8} \text{Be}  +  n ~ \rightarrow ~ ^{9} \text{Be} +  \gamma$}

\def\twobr{$ ^{9}  \text{Be} \left( \gamma,n \right) {}^{8}$Be}

\def\dgn{$^{2} \text{H} \left( \gamma,n \right) {}^{1}$H}
\def\rtwobr{$^{8} \text{Be} \left( n,\gamma \right) {}^{9}$Be}

\def\rthrbr{$\alpha \left( \alpha n,\gamma \right) {}^{9}$Be}
\def\rthrbrHEAD{$\alpha \left( \alpha n,\gamma \right) {}^{9} \text{Be}$}
\def\gn{ $\left(\gamma,n \right)$}
\def\ng{($n,\gamma$)}

\def\gray{$\gamma$-ray}

\def\higs{HI$\gamma$S}

\def\fpag{$^{19}$F($p,\alpha\gamma$)}

\def\etal{\textit{et al.}}

\def\dsumi{\displaystyle\sum\limits_{i}}
\def\dsumj{\displaystyle\sum\limits_{j}}
\def\dintfree{\displaystyle\int\limits}
\begin{document}

\preprint{APS/123-QED}
\linenumbers

\title{Cross Section Measurement of \twobr~and Implications for $\alpha + \alpha +n~\rightarrow~$\ben~in the r-Process}

\author{C.W. Arnold}
\email{arnold@lanl.gov}
\altaffiliation[Present address: ]{Los Alamos National Laboratory, P.O. Box 1663, Los Alamos, NM 87545}
\author{T.B. Clegg}%
\author{C. Iliadis}%
\author{H.J. Karwowski}%
\author{G.C. Rich}%
\author{J.R. Tompkins}%
\affiliation{Department of Physics and Astronomy, University of North Carolina at Chapel Hill, Chapel Hill, NC 27599}
\affiliation{Triangle Universities Nuclear Laboratory (TUNL), Durham, NC 27708}
\homepage{http://www.tunl.duke.edu}
\author{C.R. Howell}%
\affiliation{Department of Physics, Duke University, Durham, NC 27708}
\affiliation{Triangle Universities Nuclear Laboratory (TUNL), Durham, NC 27708}
\homepage{http://www.tunl.duke.edu}
\date{\today}

%
\begin{abstract}Models of the r-process are sensitive to the production rate of \ben~because, in explosive environments rich in neutrons, \rthrbr~is the primary mechanism for bridging the stability gaps at $A=5$ and $A=8$.   
The \rthrbr~reaction represents a two-step process, consisting of \aatobe~followed by \rtwobr.
We report here on a new absolute cross section measurement for the \twobr~reaction conducted using a highly-efficient, $^3$He-based neutron detector and nearly-monoenergetic photon beams, covering energies from E$_\gamma = 1.5$ MeV to 5.2 MeV, produced by the High Intensity $\gamma$-ray Source of Triangle Universities Nuclear Laboratory.
In the astrophysically important threshold energy region, the present cross sections are 40\% larger than those found in most previous measurements and are accurate to $\pm 10\%$ (95\% confidence).  The revised thermonuclear \rthrbr~reaction rate could have implications for the r-process in explosive environments such as Type II supernovae.
\end{abstract}

                             
\maketitle

\section{\label{sec:level1}Introduction}

The r-process likely happens in supernovae \cite{Woosley94}, in neutron star mergers \cite{neutstarmerg}, or some other high temperature ($T \geq 1$ GK) and high neutron flux ($\phi~\geq~ $10$^{20}$~cm$^{-2}$s$^{-1}$) environment.  At present, the case for an indisputable r-process site has not been made, and recent arguments point out the necessity of multiple sites \cite{RPT2007}.  Supernovae have long been cited as potential r-process factories because they produce the necessary explosive conditions and they occur frequently enough to produce substantial abundances.  The production of heavy nuclides arising from explosive nucleosynthesis at a Type II supernova site is linked to the rate of \ben~production \cite{Sas05}.

The behavior of a core-collapse supernova has been described, for example, by Woosley and Janka~\cite{woosleynature}.  
A star of 8 to 25 $M_\odot$ passes through the stages of hydrogen, helium, carbon, neon, oxygen, and silicon burning at its core, continuously growing hotter and more dense.  It eventually forms an iron-group core of Chandrasekhar mass ($\approx$1.4 $M_\odot$).  Since the core has no other source of thermonuclear energy to support the pressure, it collapses.  Energy is released in the form of neutrino radiation, while the collapse is accelerated by electron captures and photodisintegrations.  An Earth-sized iron-core collapses with a velocity of $\sim c$/4 into a single, neutron-rich nucleus about 30 km in diameter.  The collapse generates a rebounding shock wave, which ultimately stalls as it attempts to push through the in-falling matter.  The proto-neutron star briefly continues accreting matter, while radiating $\sim 10^{46}$ J in the form of neutrinos, accounting for nearly 10\% of its rest mass.  
At this stage, convection, rotation, and magnetism likely contribute to the dynamics of  the subsequent explosion.  However, self-consistent, three-dimensional models that predict explosion remain elusive \cite{Janka}. 

\subsection{\label{sec:rprocess}The r-process}

The r-process produces about half of the nuclides heavier than iron \cite{woosleynature}.  It requires a hot, neutron-dense environment where neutron captures occur so rapidly that the nucleosynthesis path is pushed far out to the neutron-rich side beyond the stability valley.  After cessation of the neutron flux, the short-lived nuclei $\beta$-decay to stable species. 

Preceding the r-process is the $\alpha$-process~\cite{Woosley92}, which is driven by charged-particle reactions and takes place when the shock has cooled from 5 GK to 3 GK over a time period of a few seconds. In this stage of nucleosynthesis a reaction path is needed to bridge the stability gaps at $A = 5$ and $A = 8$; the most efficient path is \aatobe~and \rtwobr, followed by $^9 \text{Be} + \alpha~\rightarrow~^{12} \text{C} + n$~\cite{Ter01}. As cooling continues this reaction sequence largely establishes the neutron-to-seed-nucleus ratio to which the subsequent r-process is very sensitive~\cite{Sas06}.  Too few seed nuclei will under-produce r-process nuclei, while too many seed nuclei produced in the $\alpha$-process will starve the r-process environment of neutrons.  Since r-process abundance predictions in certain stellar models are extremely sensitive to the \rthrbr~rate~\cite{Sas05, gail, mengoni}, establishing a precise rate for the formation of \ben~via the \rthrbr~reaction is required for accurately modeling nucleosynthesis in supernovae.

The \twobr~reaction 
may be used to deduce the \rtwobr~cross-section by applying the reciprocity theorem.  
Previous \ben~photodisintegration studies are numerous  \cite{old59.1,old61.1,old62.1,Berman67,Clerc68,old70.1,Hughes75,Fuji82,Barker83,Kuechler87,Gor92,Efros,Barker2000,Uts2000, Burda} but reveal relatively large cross-section uncertainties in the astrophysically important region near the neutron emission threshold. New \gn~cross-section measurements with improved accuracy are now possible using intense photon beams with small energy spreads and neutron detectors with large solid angle coverage and high efficiencies.  

In the following, Sect.~\ref{sec:exp} describes a new measurement of \twobr~and provides details of the data analysis used to obtain cross sections.  Section \ref{sec:threshbehav} describes the methods used to extract resonance parameters from the new data using proper energy dependences for partial widths of \ben~excited states. The methods employed to calculate the \rtwobr~cross section and the corresponding \rthrbr~reaction rates are described in Sect. \ref{sec:ratecalc}, along with a comparison of the determined rates with those from earlier studies. Finally, Sect. \ref{sec:discuss} presents a summary of the findings presented in this paper. 

\section{\label{sec:exp}Experiment} 
Collimated, near-monoenergetic photon~beams of 1.5 MeV $\leq E_{\gamma} \leq 5.2 ~ \mev$ were incident on a thick \ben~target located within the central bore of the neutron detector.  The absolute number of neutrons from the \gn~reaction was determined using a moderated $^{3}$He proportional counter with a high efficiency ($\sim$60\%) for detecting low-energy neutrons~\cite{mypaper}.  The absolute incident photon flux was measured using a large NaI(Tl) detector.  Photon beam energy resolution was determined with a high purity germanium (HPGe) detector.  A schematic diagram of the experimental setup is shown in Fig.~\ref{Fig:HIGSSchm}. 


\subsection{\label{sec:expmeth}Experimental setup}
\begin{figure*}[]
  \centering
  \includegraphics[width=0.95\textwidth]{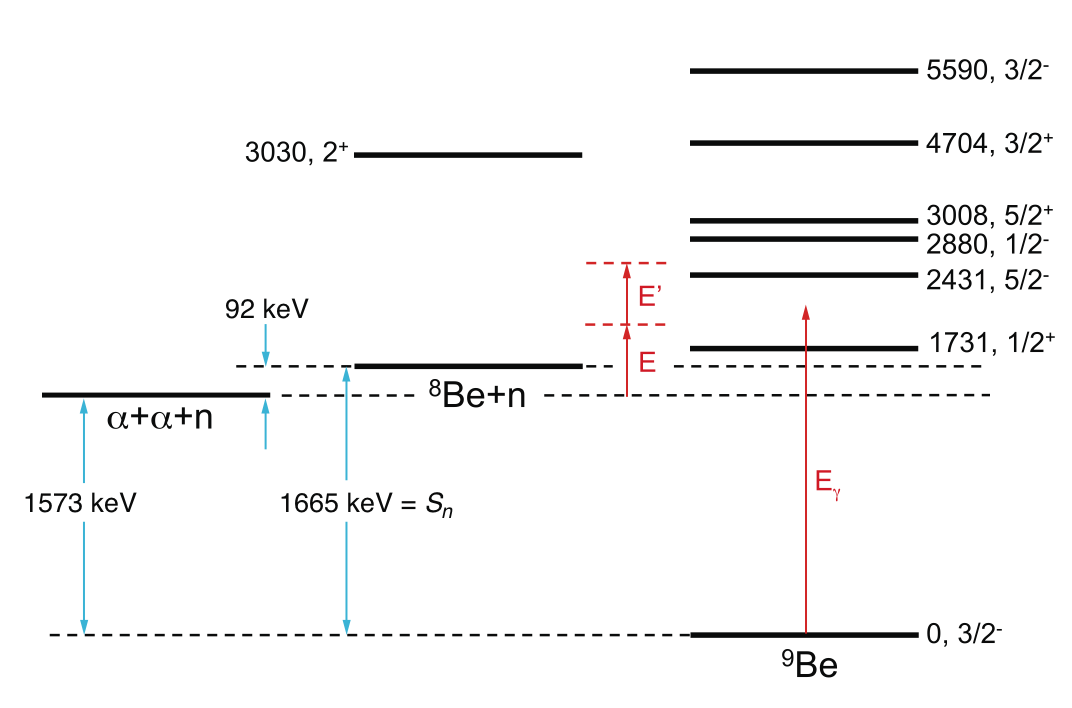}
  \caption[Level scheme relating \ben, \bee~+n, and \apa~+ n]{\label{Fig:levelscheme}(Color Online) A level scheme relating the mass energies of  \apa~+ n, \bee~+~n, and \ben.   
  Thresholds for three-body and two-body breakup of \ben~are shown to occur at incident $\gamma$-ray energies $E_\gamma$ of 1573 keV and 1665 keV, respectively. The latter is also the neutron separation energy $S_n$ for \ben.
  Energies shown for \ben~excited states are from the present work. 
  In the rate calculation (described in Sect.~\protect{\ref{sec:ratecalc}}), $E$ is the center-of-mass energy of the two \ap s.  The parameter $E'$ is the energy of the \bee~nucleus and the neutron with respect to $E$.  In this scheme, formation of \ben~at $E$ = $E'$ = 0 is very unlikely, but not prohibited because the ground state of \bee~has finite width.
 } 
\end{figure*}

Intense, collimated photon beams ($\phi \approx 3 \times 10^{7}~ $\gps) are routinely produced at the Triangle Universities Nuclear Laboratory's High Intensity Gamma Source (\higs)~by inverse-Compton scattering of free-electron-laser photons from electron bunches circulating in a storage ring~\cite{higsref}.  For the present experiment, a circularly-polarized beam was used and some flux was sacrificed to attain the high photon energy resolution needed to map the detailed behavior of the cross section at the three-body (1573 keV) and two-body (1665 keV) thresholds shown in Fig.~\ref{Fig:levelscheme}. 
Present data were taken using beam intensities of $10^{5} \leq \phi \leq 10^{6} ~$\gps~ and energy spreads of $\leq 1\%$.  

\begin{figure*}[]
\centering
\includegraphics[width=0.95\textwidth]{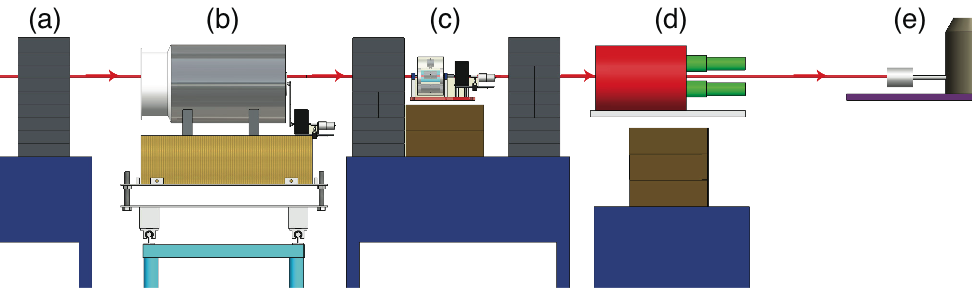}
\caption[Schematic diagram of the experimental setup.]{\label{Fig:HIGSSchm}Schematic diagram of the experimental setup for the \twobr~reaction measurements.  After collimation, the photon beam passes through scintillation paddles (not shown) and into the target room.  The photon beam then passes through the following elements: (a) ``clean-up'' collimator wall; (b) the chosen target located near the longitudinal center of the neutron counter; (c) lead attenuators located between lead collimator walls; (d) NaI(Tl) detector; (e) HPGe detector. }
\end{figure*}

The photon beam~was defined by a 12-mm diameter, 30.5-cm thick lead collimator.
It then passed successively through three thin scintillating paddles that acted as a relative photon~flux monitor, a 2.54-cm diameter hole in a lead ``clean-up'' collimator (CC), and $\sim$1.5 m of air before reaching a second CC (shown as (a) in Fig.~\ref{Fig:HIGSSchm}) placed directly in front of the neutron detector.

\begin{table*}
	\begin{center}
	\caption[Table of target properties.]{\label{tab:targetsandprop} Targets used in the present experiment and their physical properties.}
	\begin{tabular}{ ccccc }
		\hline\hline 
		\multirow{2}{*}{Material}   & Length  & Density        & Mass  & Thickness \\ 
		 & (cm)            &   (g cm$^{-3}$) & (g mol$^{-1}$)      & (nuclei cm$^{-2}$) \\
		\hline
		\ben  & 2.54 & 1.848 & 9.012  &3.14$\times$10$^{23}$\\
		Graphite & 2.54 & 1.700 & 12.01  & 2.89$\times$10$^{23}$\\
		\dtwoO  & 7.59 & 1.106 &  20.04  & 5.05$\times$10$^{23}$\\
		\hline\hline
	\end{tabular}
	\end{center}
\end{table*}

At each energy, photons~impinged successively on one of three thick, 19-mm diameter cylindrical targets (\ben,~\dtwoO, graphite) described in Table~\ref{tab:targetsandprop}, or air, as they passed through the neutron detector.  The heavy water target was bombarded under the same experimental conditions as the \ben~target, allowing for normalization of the \twobr~measurements to the well-known \dgn~cross section \cite{Sch05} and for calibration of the neutron detector efficiency.  The graphite target was used to determine beam-induced backgrounds in the neutron detector. 

\begin{figure}[]
\centering
\includegraphics[width=0.5\textwidth]{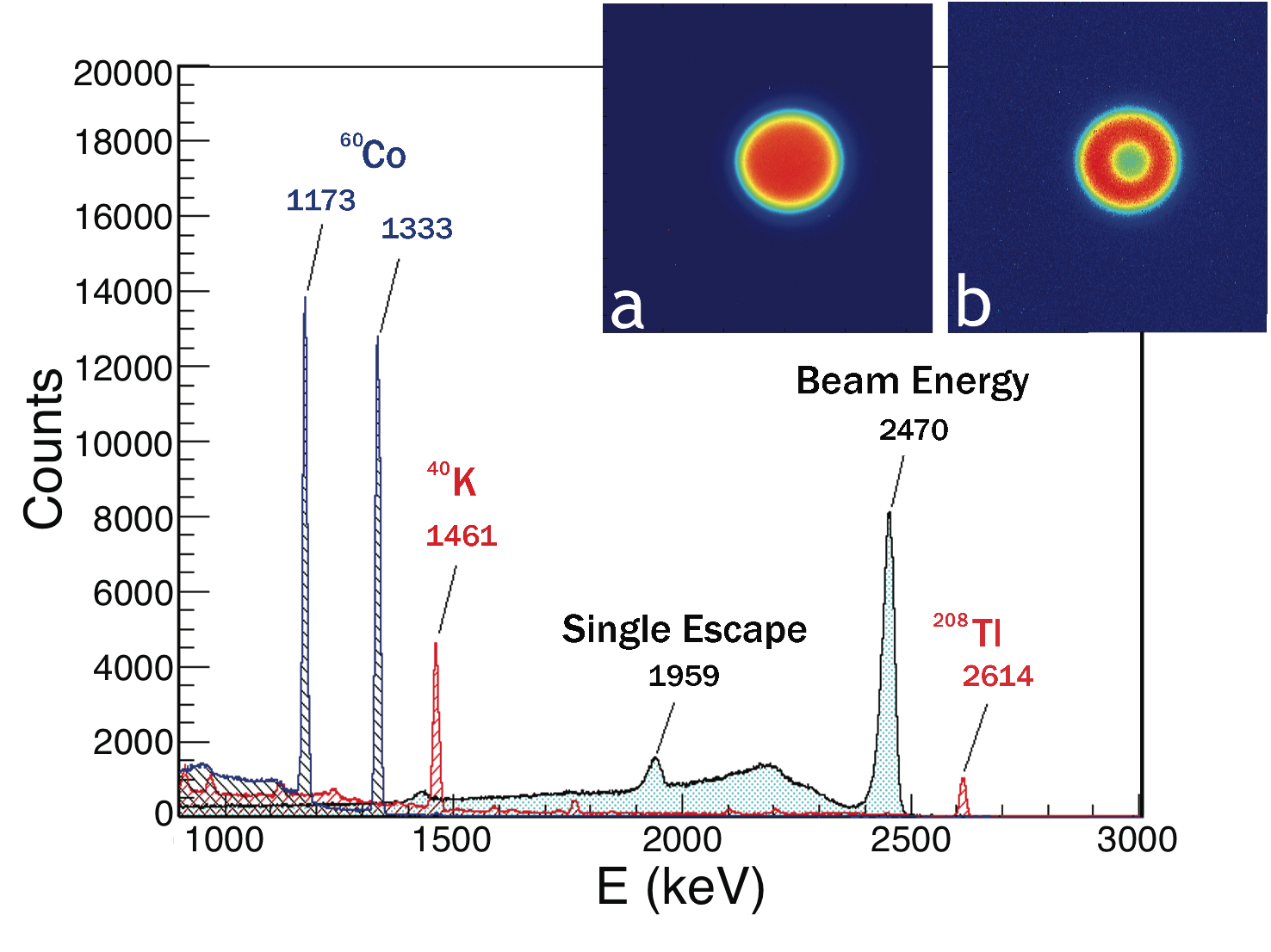}
\caption[A typical HPGe spectrum.]{(Color Online) A HPGe spectrum for 2470-keV photon beam with a resolution of $\Delta$E/E = 1\%.  Spectra obtained with $^{60}$Co, $^{40}$K, and $^{208}$Tl calibration sources are overlaid.  Images showing confirmation of target alignment are inset. (a) An unattenuated beam profile.  Flux was nearly constant across the 12-mm diameter and decreased rapidly at the edges. (b) Contrast from a 4-mm diameter lead alignment pellet confirmed the axial placement of the target.}
\label{Fig:g_stats}
\end{figure}

To increase the efficiency of data collection, targets were remotely rotated into the beam using a four-position Geneva mechanism, which also assured reproducible alignment of the axis of each cylindrical target sample with the beam axis. Axial alignment of the target was confirmed using an alignment pellet and a photon~beam imaging system \cite{Sun08} (see Fig.~\ref{Fig:g_stats}).

Downstream of the target,  
lead of various thicknesses could be inserted to attenuate the beam by up to a factor of 100. 
Further downstream, the remaining photons~were incident on either a NaI(Tl) detector 
 or a HPGe detector, depending on whether photon flux or energy was being measured.  The lead attenuators facilitated simultaneous high neutron counting rates and negligible NaI(Tl) signal pile-up.  Data acquisition dead-times were assured to be small. The counts measured in the NaI(Tl) detector were corrected for detection efficiency as well as for attenuation through the lead and targets.  

\subsection{\label{detectors}Detector calibration}
Absolute measurements of the number of photons~on target and the number of emitted neutrons from the reaction were needed to determine the total cross section of \twobr. Thus, it was essential to determine the absolute energy-dependent detection efficiencies of the neutron detector and the large NaI(Tl) detector.
The active neutron detection elements were 18 tubular proportional counters, each containing $^{3} \text{He}$ at $ 6.1\times 10^{5} ~ \text{Pa}$.  The tubes, embedded in a cylindrical polyethylene body that served as a neutron moderator, were arranged in concentric inner ($I$) and outer ($O$) rings of nine equally spaced detectors each.
The energy-dependent efficiency of the neutron counter was determined in an extensive study~\cite{mypaper}. 
The ratio of counts in the inner and outer rings ($I/O$ ratio) provided a coarse estimate of the average neutron energy. 

The total efficiency of the large NaI(Tl) detector was found to be nearly constant ($98.3 \pm 1.7\%$) over the experimental energy range using the Monte-Carlo codes \protect{\sc{geant4}} and  \protect{\sc{mcnpx}}.  The results were consistent with data obtained using the \fpag~reaction, taken using a mini-tandem accelerator \cite{minitandem}. This measurement provided a determination of the absolute detection efficiency for 6.13 MeV photons~\cite{mythesis, TPR}.
\begin{figure}[]
\centering
\includegraphics[width=0.5\textwidth]{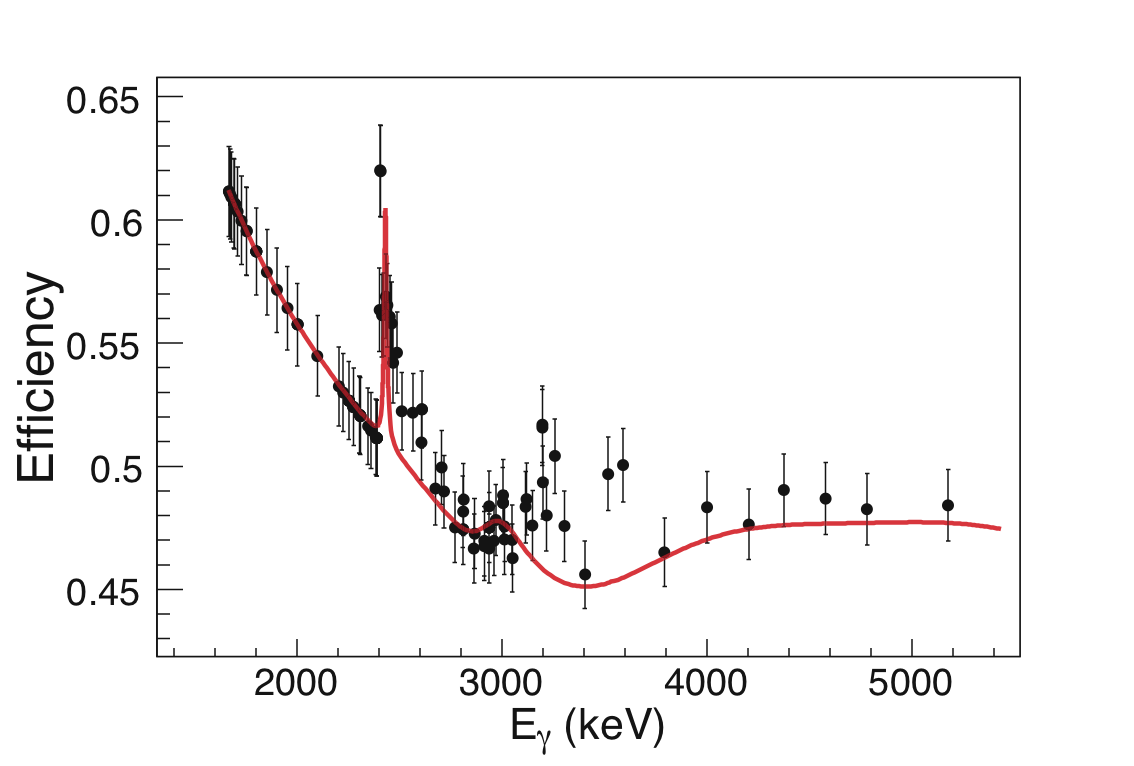}
\caption[]{\label{Fig:be9eff}(Color Online) Effective neutron detector efficiency versus photon~energy for \twobr~with error bars representing the 4.6\% systematic uncertainty.  Above $E_\gamma = 2431$~keV, average neutron energy, and thus detection efficiency, was determined using the experimental $I/O$ ratio. The (red) curve is modeled using Eq.~\ref{eq:effeq} with the branching ratios listed in Table \ref{tab:results}.  Discrepancies in the 3 to 4 MeV range are not well understood, but do not affect the present calculation for the astrophysical \rthrbr~reaction rate discussed in Sect.~\protect{\ref{sec:ratecalc}}.}
\end{figure}

While the efficiency of the large NaI(Tl) detector was constant for photons within the experimental energy range, the effective efficiency of the neutron detector for photoneutrons from \ben, $\epsilon_\textrm{eff} \left( E_\gamma \right)$, varied and is shown in Fig. \ref{Fig:be9eff}.  As can be seen in Fig. \ref{Fig:levelscheme}, the average energy for neutrons which decay to the ground state of \bee~is $E_{\gamma} - S_{n}$, and for these neutrons the detection efficiency $\epsilon_n$ may be described using a sixth-order polynomial in $E_n$.  Above $E_\gamma = 2431$~keV, multiple decay channels exist and the average neutron energy is no longer proportional to the photon energy. Instead, it becomes nonlinear and strongly dependent on the photon energy, which causes $\epsilon_\textrm{eff}$ to deviate from the polynomial that describes $\epsilon_n$.  To account for this, 
the effective efficiency was described using
\begin{eqnarray}
	\epsilon_\textrm{eff} \left( E_\gamma \right) = \epsilon_n \left(E_{\gamma} - S_{n} \right) \left(\dsumj{\frac{\beta_{j}\sigma_{j}}{\sigma_{tot}}} \right) + \nonumber\\ 
	\epsilon_n \left( \delta E_{n} \right) \left(\dsumj{\frac{ \left(1-\beta_{j} \right) \sigma_{j}}{\sigma_{tot}}} \right),
\label{eq:effeq}
\end{eqnarray}
where for level $j$, $\sigma_{j}$ is the contribution from state $j$ to the total cross section $\sigma_{tot}$, and $\beta_{j}$ is a branching ratio for state $j$ to the ground state of \bee.  This form assumes that a newly opened neutron branch will decay with a small neutron energy $\delta E_{n}$.  The detection efficiency for the fraction of the total cross-section decaying to the ground state is described by the simple polynomial $\epsilon_n \left(E_{\gamma} - S_{n} \right)$, while the efficiency of new branches is $\epsilon_n \left( \delta E_{n} \right) \approx \epsilon_{max}$.  This model is compared in Fig.~\ref{Fig:be9eff} to a point-by-point determination of the neutron detector efficiency constructed using the $I/O$ ratio.  The branching ratio for the $5/2^{-}$ state at 2431 keV was taken from Ref. \cite{CandC}.  Other branching ratios were chosen to make the model congruent with the point-by-point analysis.  The contribution of states other than the 1/2$^{+}$ state to the \rthrbr~rate will be shown in Sect.~\protect{\ref{sec:ratecalc}} to be nearly independent of the choice of branching ratios.

\subsection{\label{sec:Analyses}Data analysis}
We performed two analyses of the yield data: (i) assuming a monoenergetic photon beam; and (ii) assuming a photon beam with a finite energy width, requiring deconvolution of the photon beam energy profile to interpret the neutron yield accurately. 

Under the assumption of a monoenergetic beam, the cross section may be written as
\begin{equation}
\sigma = \frac{N_{n}}{N_{\gamma}\cdot(N_{T}/A)\cdot\epsilon_{n}},
\label{Eq:MBA}
\end{equation}
where $N_n$ is the number of detected neutrons, $N_\gamma$ is the number of incident photons, $N_{T}/A$ is the effective number of target nuclei per unit area, and $\epsilon_{n}$ is the neutron detector efficiency. The quantity $N_T/A$ was determined by comparing neutron yields from target-in and target-out runs.  Thick targets required a photon~energy-dependent correction of the form
\begin{equation}
\eta = \frac{1-e^ {-\mu t  }}{\mu t}
\label{Eq:TTC},
\end{equation}
where $\mu$ is a material specific attenuation coefficient and $t$ is the thickness of the target. This correction accounted for the reduction in the number of incident photons caused by interactions within the target volume.

%
 Total \twobr~cross-section uncertainties for $\sigma$ in Eq.~\ref{Eq:MBA} were found to be 3.2\% (statistical) and 4.6\% (systematic).  The largest contributions to the uncertainties came from the absolute efficiencies of the neutron and NaI(Tl) detectors and the exact photon beam flux loss associated with lead attenuators. Numerical values for cross sections and a detailed analysis of the associated experimental uncertainties  are available in Ref.~\cite{mythesis}.

\begin{figure*}
  \centering
  \includegraphics[width=.95\textwidth]{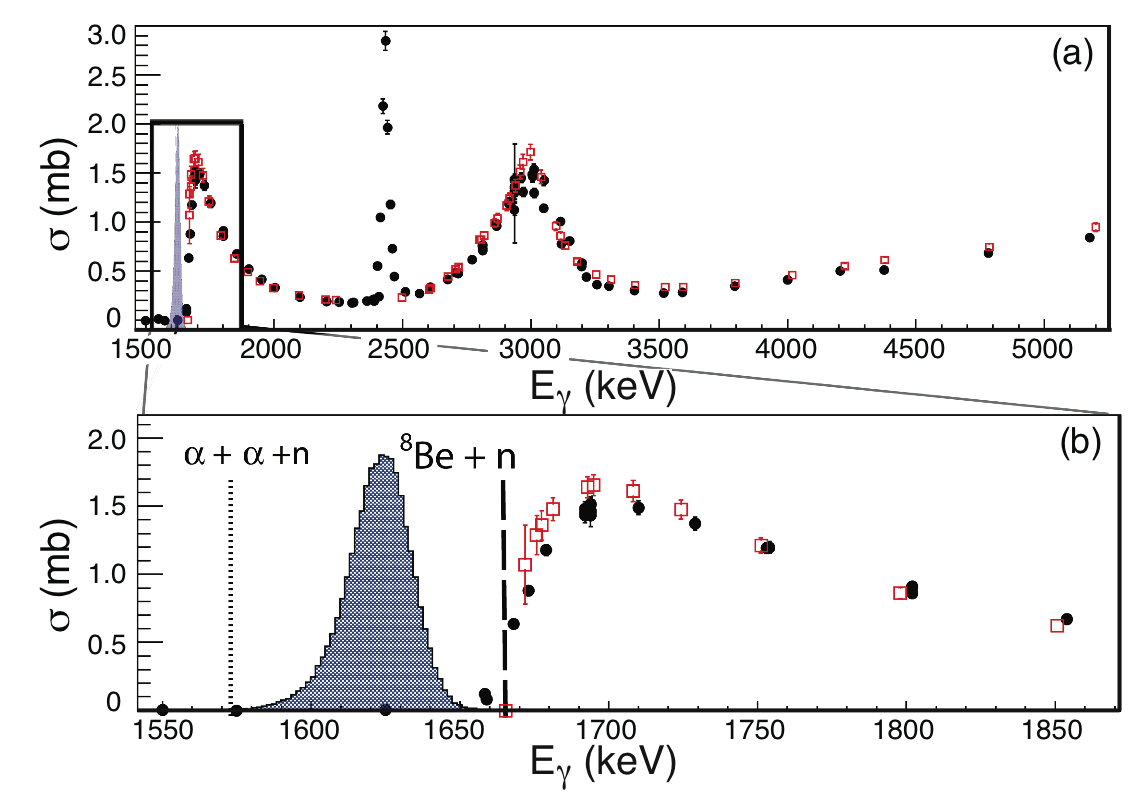}
  \caption{\label{fig:bothcs}(Color Online) A plot of monoenergetic beam cross sections (black dots) and deconvoluted cross sections (open red squares). Error bars are shown when uncertainties are larger than the points. The uncertainty shown for the monoenergetic beam data is purely statistical, while that shown for the deconvoluted data also includes statistical and systematic flux uncertainties associated with energy binning near threshold. (a) The full experimental range.  The resonance at 2431 keV was too narrow for accurate deconvolution. Several points with high uncertainty, resulting from low-statistics runs, are apparent at $\sim$2950 keV; these data, taken at the same energy, were combined for the purpose of deconvolution. (b) An expanded view of the boxed region near threshold better shows the differences between the two methods. The dotted (dashed) vertical line denotes the three-body (two-body) threshold. The sample photon beam profile shown (blue) is peaked at 1625 keV with an energy spread which was typical of this experiment.}
\end{figure*}

The second analytic approach treated the reality that the photon beam was not truly monoenergetic. Such a treatment is especially important near threshold, where the cross section changed significantly within the energy spread of the beam (see Fig. \ref{fig:bothcs}). 
The experimental yield $Y$ may be defined as
\begin{equation}
Y = \frac{\int{f  \cdot\sigma_{t}\cdot\epsilon_{n}\cdot (N_{T}/A)dE_{\gamma}}}{\int{f  }dE_{\gamma}} = \frac{N_{n}}{N_{\gamma}}
\label{eq:yield},
\end{equation}
where $f $ is an energy-dependent function describing the energy distribution of the photon~beam and $\sigma_{t}$ is a trial cross section.  

To determine $f$, the detector response function was deconvolved from the HPGe spectrum at each beam energy. The resulting spectra were then normalized such that
\begin{equation}
\int{f dE_\gamma}= N_{\gamma}
\label{intgammas},
\end{equation}
where $N_{\gamma}$ was determined using the NaI(Tl) detector.  
The trial cross section $\sigma_t$ was assumed to be the sum of six Breit-Wigner equations (BWEs), each of which had three free parameters corresponding to the resonance energy $E_{R}$, the neutron partial width $\Gamma_{n}$, and the transition strength B(E1) or B(M1) of an excited state in \ben.

\begin{figure}
  \centering
  \includegraphics[width=.4\textwidth]{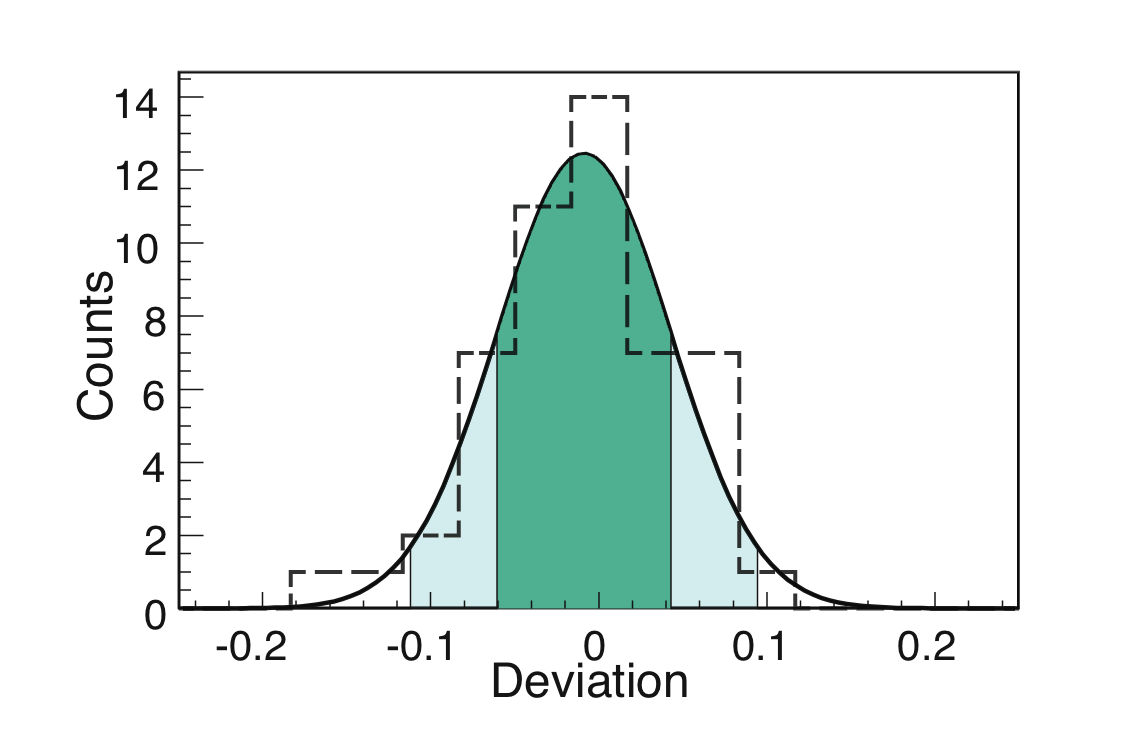}
  \caption{\label{fig:errhist}(Color Online) Histogram of the relative deviation between $Y$ and $Y^{*}$.  The Gaussian fit to the error is centered just below zero.  The $\pm 1\sigma$ interval is darkly shaded (teal) and the $\pm 2\sigma$ interval is shaded lightly (blue).}
\end{figure}

Histograms were constructed for the other components of $Y (\sigma, \epsilon_{n}, N_{t}/A$) as a function of photon~energy. Each bin of the histogram then represented the respective component of the yield over $dE$, the width of the bin.  Thus, the calculated yield $Y^*$was given by
\begin{equation}
Y^{*} = \frac{\dsumi f_{i}\cdot\sigma_{i}\cdot{\epsilon_{n}}_{i}(N_{t}/A)_{i}}{\dsumi{f_{i}}},
\end{equation}
where the width of the $i^{\text{th}}$ bin was $\sim$1.6 keV.
In this way the yield-weighted \textit{effective} energy, $E_\gamma^*$ was defined as,
\begin{equation}
E_\gamma^{*} = \frac{\dsumi{E_{\gamma}}_{i}f_{i}\cdot\sigma_{i}\cdot{\epsilon_{n}}_{i}(N_{t}/A)_{i}}{\dsumi f_{i}\cdot\sigma_{i}\cdot{\epsilon_{n}}_{i}(N_{t}/A)_{i}}.
\end{equation}
The trial cross section was then iteratively adjusted (over $\sim 8$ steps) until the global deviation between the calculated yield and the experimental yield was minimized.  This process resulted in a deconvoluted cross section shown in Fig. \ref{fig:bothcs} along with that deduced from the monoenergetic beam analysis.  Figure \ref{fig:errhist} shows a histogram of the deviations between experimental yields and yields calculated from the deconvoluted cross section for 52 data points.  The relative error is within $\pm 5\%$ at the 68\% confidence interval and $\pm 10\%$ at the 95\% confidence interval.
\subsection{\label{sec:TBCS}The \twobr~cross section}

\begin{figure*}[!ht]
\centering
\includegraphics[width=0.95\textwidth]{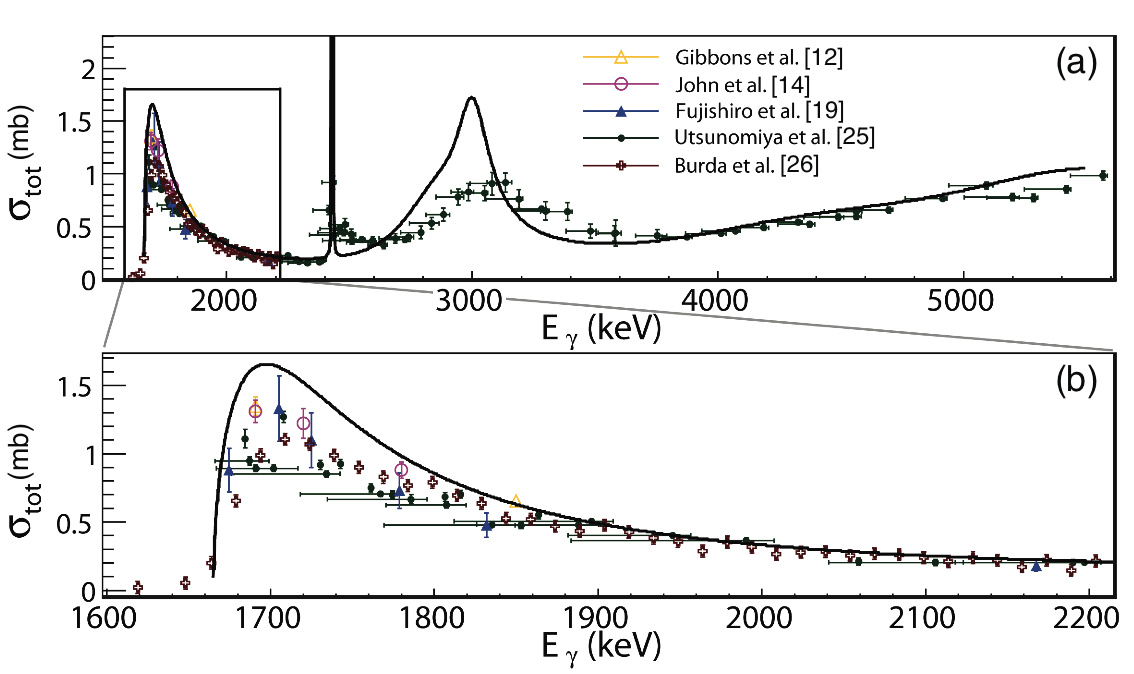}
\caption[World data for \twobr~at threshold]{\label{fig:WD_Thresh}
(Color Online) 
(a) Data for the total \twobr~cross section collected using several different $\gamma$-ray sources, including virtual photons from inelastic electron scattering \protect{\cite{Burda}} and real photons from both inverse Compton scattering \protect{\cite{Uts2000}} and natural radioisotopes \protect{\cite{old59.1,old62.1,Fuji82}}.
(b) An expanded view of the boxed region better shows the different evaluations near the threshold.}
\end{figure*}

From the two-body threshold energy of $E_{Th} =  S_n = 1665$ keV to 1900 keV, the present cross sections are larger than most of the previously reported data. 
Notice in Fig. \ref{fig:bothcs} that deconvolution changes the shape of the cross section at the threshold, transferring strength from below to above $E_{Th}$.  
The need for deconvolution could well explain some disagreement in this region with yields measured in earlier experiments using photon beams having larger energy spreads than those used in the present experiment~\cite{Uts2000, Gor92}.

The narrow 5/2$^{-}$ state at 2431 keV was far better resolved in the present experiment than in previous works.  The present experimental yield peaks more than a factor of 3 above the peak measured by Ref.~\cite{Uts2000} and nearly a factor of 2 above that of Ref.~\cite{Gor92}. 
The present data, over the broad 1/2$^{-}$ and 5/2$^{+}$ resonances near 3 MeV, are in fair agreement with the data of Refs.~\cite{Hughes75,Gor92} but not in agreement with the results of Ref.~\cite{Uts2000} which were obtained with a large photon beam energy spread. 

At energies above the broad peak at 3 MeV the present data agree with Refs. \cite{Uts2000,Hughes75}, but not with Ref. \cite{Gor92}.  A 3/2$^{+}$ state at 4.7 MeV and a 3/2$^{-}$ state at 5.6 MeV are the next known  excited states in \ben~\cite{narrowstate}.
These are broad states~\cite{Dixit} which decay more strongly through the 2$^{+}$ excited state in \bee. 

\section{\label{sec:threshbehav}Threshold Behavior and Energy Dependence} 
Previous measurements of the \emph{direct} $^9$Be$+\gamma \rightarrow \alpha + \alpha + n$~three-body reaction in the energy range 1570 keV $ < E_{\gamma}< 1670$ keV have yielded only an upper limit to the total cross section of 93 nb \cite{threebody}, and the present experiment was not sufficiently sensitive to improve this.  Above the two-body threshold at 1665 keV, the total $^9$Be($\gamma$,n) cross section rises rapidly to exceed 1 mb. This feature is most logically attributed to the newly-opened decay channel, and thus excitation of the broad 1/2$^+$ state located immediately above the two-body threshold is most frequently followed by decay to $^{8} \text{Be}_{gs} + n$.

\subsection{\label{sec:BNT}Behavior near threshold}
Any resonance close to threshold experiences a distortion of its normal Lorentzian cross-section shape.
When a level of spin $J$ is isolated from other levels of the same spin and parity, a one-level R-matrix approximation may be used to describe the contribution of this level to the \gn~cross section.  For \gn~reactions, this takes the form of the BWE for an isolated resonance \cite{Iliadisbook}:
\begin{equation} \label{eq:BWE}
	\sigma_{\gamma,n}(E_{\gamma}) = \frac{\pi}{k_{\gamma}^{2}}\frac{2J +1}{2(2I+1)}\frac{\Gamma_{\gamma}\Gamma_{n}}{(E_{\gamma}- E_{R})^{2} + \frac{1}{4}\Gamma^{2}},
\end{equation}
with $I$ the spin of the target nucleus and $k_\gamma^2$ given by 
\begin{equation} \label{eq:ksubgammasquared}
	 k_{\gamma}^{2} = \left(\frac{E_{\gamma}}{\hbar c}\right)^{2}.
\end{equation}
The neutron partial width $\Gamma_{n}$ is generally written as \cite{blattandweiss}
\begin{equation} \label{eq:genwidth}
	\Gamma_{n} = 2\gamma^{2}P_{\ell},
\end{equation}
where $\gamma^{2}$ is the reduced width and $P_{\ell}$ is the penetration factor.  The reduced width incorporates the unknown parts of the nuclear interior while $P_{\ell}$ is completely determined by the conditions outside the nucleus and may be written as
\begin{equation}
P_{\ell}=R\left(\frac{k}{F^{2}_{\ell}+G^{2}_{\ell}}\right),
\end{equation}
\noindent where $R$ is the channel radius,  $k$ is the wave number, and $\ell$ is the neutron orbital angular momentum \cite{Iliadisbook}. The channel radius $R$ is defined as \cite{Iliadisbook}
\begin{equation} \label{eq:channelRadius}
	R = r_{0}(A_{t}^{1/3}+A_{p}^{1/3}),
\end{equation}
\noindent with $A_t$ and $A_p$ the mass numbers of the target and projectile, respectively, and $r_0 = 1.44$~fm. 
 For neutrons, the Coulomb wave functions, $F_{\ell}$ and $G_{\ell}$, are related to spherical Bessel ($j_{\ell}$) and Neumann ($n_{\ell}$) functions by $F_{\ell} = (kr) j_{\ell} \left(kr \right)$ and $G_{\ell} = (kr) n_{\ell} \left(kr \right)$.
In the cases of \gn~and~\ng~reactions, the penetration factors can be written analytically, and for $0 \leq \ell \leq 3$, P$_{\ell}$ is~\cite{blattandweiss}
\begin{subequations}
	\begin{eqnarray}
		P_{0} = &kR = \sqrt{\xi E_{n}} \label{eq:pzero}, \\
		\nonumber\\
		P_{1} = &\dfrac{(\xi E_{n})^{3/2}}{1+\xi E_{n}} \label{eq:pone}, \\
		\nonumber\\
		P_{2} = &\dfrac{(\xi E_{n})^{5/2}}{9+3\xi E_{n}+(\xi E_{n})^{2}} \label{eq:ptwo}, \\
		\nonumber\\
		P_{3} = &\dfrac{(\xi E_{n})^{7/2}}{225+45\xi E_{n}+6(\xi E_{n})^{2}+(\xi E_{n})^{3}} \label{eq:pthree}.
	\end{eqnarray}
\end{subequations}
\noindent In these expressions $R$ is again the channel radius defined in Eq. \ref{eq:channelRadius}; the neutron energy is related to the photon energy by $E_{n} = E_{\gamma} - S_{n}$; and we define $\xi \equiv 2\mu R^{2} \hbar^{-2}$, where $\mu$ is the reduced mass of $^8 \text{Be} + n$.

\begin{figure}[]
  \centering
  \includegraphics[width=0.5\textwidth]{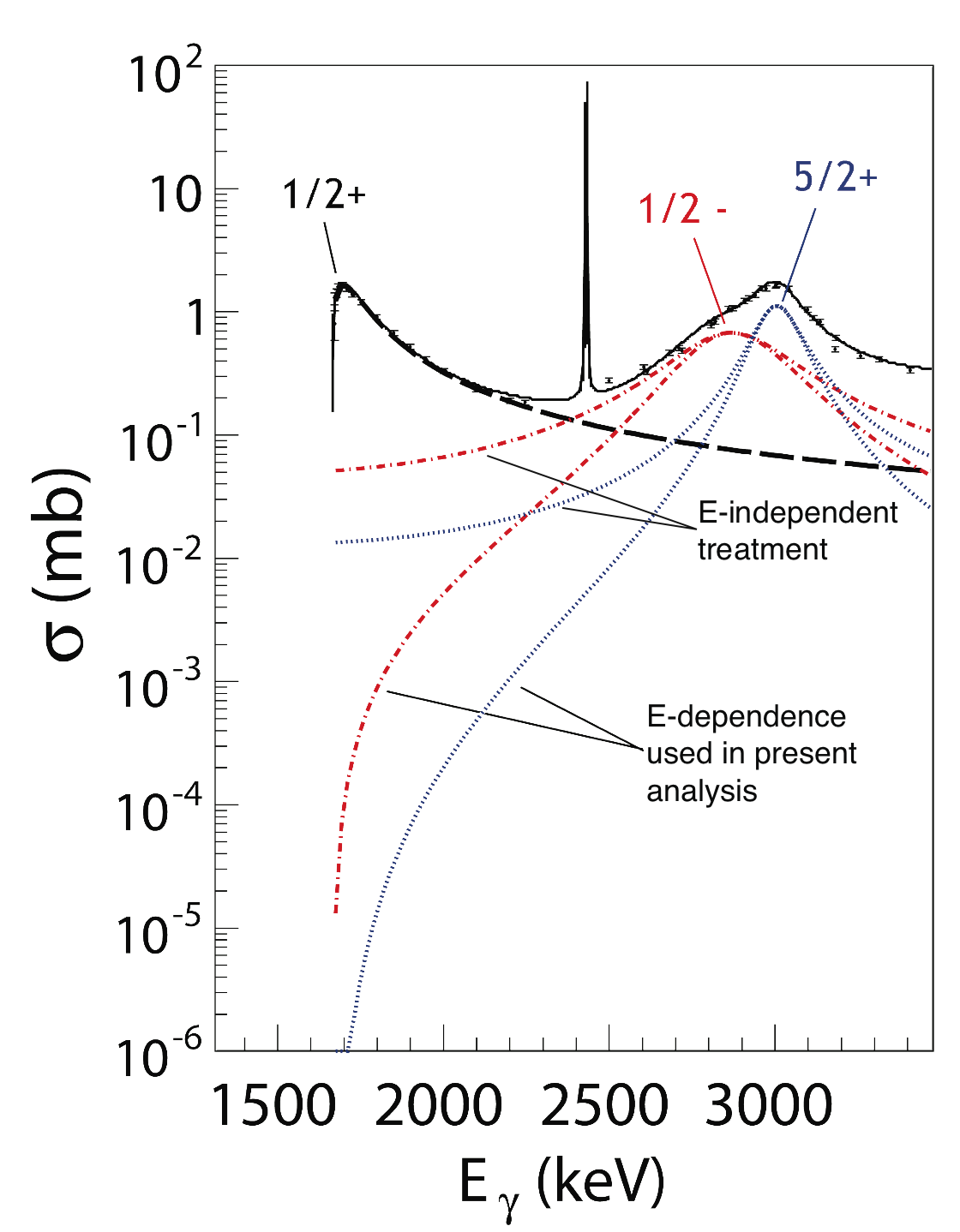}
  \caption{\label{Fig:thresh_behav}(Color Online) Example of the differences between energy-independent and energy-dependent calculations of the \twobr~cross sections at threshold.  The dashed (black), dot-dashed (red) and dotted (blue) lines show the 1/2$^{+}$, 1/2$^{-}$, and 5/2$^{+}$ states, respectively.  Energy-dependent cross-section contributions from the 1/2$^{-}$ and 5/2$^{+}$ states are identified and contrasted with energy-independent determinations of these states; at threshold, the latter are orders of magnitude too large.  }
\end{figure}

One must include $P_{\ell}$ energy-dependence for the 1/2$^{+}$ threshold resonance to obtain a good fit to the \twobr~cross-section data.  However, previous works have not included energy dependence in the tails of the broad, higher-lying states in \ben.  The result of this incomplete treatment has been to inflate the previously-deduced off-resonance contributions to the \rthrbr~rate by as much as a factor of five.  
The states in \ben~excited by an $L=1$ photon are coupled to the ground state of \bee~through emission of a neutron with a specific orbital angular momentum $\ell$ determined by the spin and parity of the excited state. 
For the excited states considered in this paper, those with $J^\pi = 1/2^+$, 1/2$^-$, 3/2$^+$, 3/2$^-$, 5/2$^+$, and 5/2$^-$, decay to the ground state of \bee~through emission of a neutron with $\ell = 0$, 1, 2, 1, 2, and 3, respectively.
The value of $\ell$ determines the form of $P_{\ell}$ (Eqs. \ref{eq:pzero} - \ref{eq:pthree}), which in turn determines the energy dependence of the neutron partial width (defined in Eq. \ref{eq:genwidth}) for the excited state in~\ben, and ultimately the behavior of its cross section near the threshold. Figure \ref{Fig:thresh_behav} displays the relative contributions of the 1/2$^{+}$, 1/2$^{-}$, and 5/2$^{+}$ resonances to the total cross section. The latter two are shown with and without the proper P$_{\ell}$ energy dependence.


The energy-dependent \gray~partial widths may be cast in terms of reduced transition probabilities \cite{blattandweiss}. For E1 and M1 transitions one finds
\begin{equation}
\Gamma_{\gamma} (E1) = \frac{16\pi}{9}\alpha(\hbar c)^{-2}E_{\gamma}^{3}B(E1)\downarrow
\label{Eq:BE1},
\end{equation}
 \noindent and
\begin{equation}
\Gamma_{\gamma} (M1) = \frac{16\pi}{9}\alpha(2M_{p}c^{2})^{-2}E_{\gamma}^{3}B(M1)\downarrow
\label{Eq:BM1},
\end{equation}
\noindent where $\alpha$ is the fine structure constant. Note that the strength of a transition from the ground state to an excited state (B$\uparrow$) is related to the strength of transition from that excited state to the ground state (B$\downarrow$) by
\begin{equation}
	B \! \uparrow = \frac{2J_{x}+1}{2J_{0}+1}B \! \downarrow.
	\label{eq:uptodown}
\end{equation}
\noindent Transitions from the ground state of \ben, $J_{0} = 3/2^{-}$, to an excited state, $J_{x} = 1/2$, 3/2, or 5/2, yield ratios of 0.5, 1, and 1.5,  respectively, in Eq.~\ref{eq:uptodown}.  Since $B\downarrow$ are the strengths of transitions used in the calculation of the \rthrbr~rate, they will be used exclusively in the following discussion.  For each resonance, the reduced width $\gamma^2$, the transition strength B(E1)$\downarrow$ or B(M1)$\downarrow$, and the resonance energy $E_{R}$ are determined by fitting the data. Table \ref{tab:results} displays the parameters determined for each resonance.  
The sub-eV width of the narrow 5/2$^-$ state precluded experimental determination of the associated $\Gamma_n$, and therefore the width reported in Ref. \cite{narrowstate} was adopted for the present rate calculation.
Known contributions from B(E2)$\downarrow$ for the negative parity states were negligible compared to B(M1)$\downarrow$~\cite{narrowstate}.

\begin{table*}
	\begin{center}
	\caption{ \label{tab:results} Resonance parameters and neutron branching ratios from the present work.  The latter were determined from the neutron detector efficiency analysis discussed in Sect.~\protect{\ref{detectors}}}
	\begin{threeparttable}
	\begin{tabular*}{\textwidth}{@{\extracolsep{\fill}} ccccccc} 
		\hline\hline
  		\multirow{3}{*}{ $J^{\pi}$ } 	&	\multirow{3}{*} {$\chi_{\lambda}$}	&	 \multirow{3}{*} { $E_{R} \left(\mev \right)$ }	&	 B$\left( \chi_\lambda \right)$ &		\multirow{3}{*} { $\Gamma_{\gamma} \left( \text{eV} \right) $ }	&	\multirow{3}{*}{ $\Gamma_{n} \left( \text{keV} \right) $ }	&	\multirow{3}{*}{ $\beta_{j} \left( \% \right) $ }\\
		& 	&	&	E1$\rightarrow \left( \text{e}^{2} \text{fm}^{2} \right)$	& 	 &	&	 \\
		&	&	&	M1 $\rightarrow \left( \mu_N^{2} \right)$ &	&	&	\\
		\hline
		1/2$^{+}$ & E1 &   $1.731 \pm 0.002$ &  $0.136 \pm 0.002$ &  $0.738 \pm 0.002$ &  $213 \pm 6$       &  100\\
		5/2$^{-}$ & M1 &   $2.431 \pm 0.004$ &  $0.587 \pm 0.027$ &  $0.098 \pm 0.004$ &   $0.77$\tnote{a,b} &  6\tnote{c}\\
		1/2$^{-}$ & M1 &   $2.880 \pm 0.016$ & $6.5 \pm 0.7$    &  $1.8 \pm 0.2$     &  $393 \pm 18$      &  100\\
		5/2$^{+}$ & E1 &  $3.008 \pm 0.004$ &  $0.016 \pm 0.002$ &  $0.45 \pm 0.07$   &  $163 \pm 15$      &  70\\
		3/2$^{+}$ & E1 &   $4.704$\tnote{b}    &  $0.068 \pm 0.007$ &  $7.8 \pm 0.4$     &  $1541 \pm 115$    &  38\\
		3/2$^{-}$ & M1 &  $5.59$\tnote{b}     &  $7.8 \pm 2.1$  &  $15.7 \pm 4.2$    &  $941 \pm 164$     &  38\\
		\hline\hline
	\end{tabular*}
	\begin{tablenotes}
	\item[a] \scriptsize{This value could not be obtained using the present data.}
	\item[b]\scriptsize{This value was fixed in accordance with Ref.~\protect{\cite{narrowstate}}.}
	\item[c] \scriptsize{This was fixed in accordance with Ref.~\protect{\cite{CandC}}.}
	\end{tablenotes}
	\end{threeparttable}
	\end{center}
\end{table*}

\subsection{Narrow resonance treatment}
In the case of narrow resonances, $\Gamma_{\gamma}$ may be deduced by integrating the cross section.  
For this to be valid: (a) the resonance must be sufficiently isolated from other resonances, (b) the neutron and \gray~partial widths must be small enough to be considered energy-independent, and (c) the neutron partial width must be much larger than the \gray~partial width such that $\Gamma_{n} \approx \Gamma$.  
With these three conditions satisfied, Eq.~\ref{eq:BWE} may be integrated for the 5/2$^{-}$ resonance, yielding
\begin{equation}
	\int\sigma_{\gamma,n}(E_{\gamma})dE_{\gamma} = \frac{3}{2}\left(\frac{\hbar c\pi}{E_{R}}\right)^{2}
	\Gamma_{\gamma}.
	\label{eq:integratedBWF}
\end{equation}
The value obtained from this analysis is shown in Table \ref{tab:results}.

\subsection{\label{sec:ResParams}The 1/2$^{+}$ threshold resonance}
As will be shown in Sect.~\ref{sec:ratecalc}, the 1/2$^{+}$ threshold resonance is \emph{the} largest contributor to the \rthrbr~reaction rate.
Table~\ref{tab:onehalfparams} contains resonance parameters for this state from several works.
\begin{table*}
\caption[]{ Resonance parameters and reduced transition probabilities of the $1/2^{+}$ state of \ben~from virtual and real photon experiments. Refs. \cite{Barker2000, Barker83, Sumi} include reanalyses of data originally analyzed by Refs. \cite{Kuechler87, Fuji82, Uts2000}, respectively. }
\begin{center}
\begin{tabular}{ cccccc }
\hline\hline
Reaction  & Ref. &  $E_{R}$~(MeV) & $\Gamma_{n}$~(keV) & $\Gamma_{\gamma}$~(eV) & B(E1)$\downarrow$~(e$^{2}$fm$^{2}$)\\
\hline
 ($e,e'$)      & Clerc \etal \protect{\cite{Clerc68}} & 1.78  & $ 150 \pm 50 $ & 0.3    & $ 0.050 \pm 0.020 $ \\
 ($e,e'$)      & Kuechler  \etal \protect{\cite{Kuechler87}} & 1.684 & $ 217 \pm 10 $ & 0.27     & 0.054 \\
 ($e,e'$)      & Glick  \etal \protect{\cite{Gli91}} & 1.68  & $ 200 \pm 20 $& 0.34     & 0.068 \\
 ($e,e'$)      & Barker  \etal \protect{\cite{Barker2000}} & 1.732 & 270       & 0.75     & 0.137 \\
 ($e,e'$)      & Burda  \etal \protect{\cite{Burda}} & 1.748 & $ 274 \pm 8 $ & $ 0.302 \pm 0.045$ & 0.054 \\
 \hline
 ($\gamma,n$) & Barker \etal \protect{\cite{Barker83}} & 1.733 & $ 227 \pm 50 $ & 0.577     & $ 0.106 \pm 0.018$ \\
 ($\gamma,n$) & Angulo \etal \protect{\cite{NACRE}} & 1.731 & $ 227 \pm 15 $ & $ 0.51 \pm 0.10 $ & $ 0.094 \pm 0.020$ \\
 ($\gamma,n$) & Utsunomiya  \etal \protect{\cite{Uts2000}} & 1.748 & $ 283 \pm 42 $ & 0.598     & $ 0.107 \pm 0.007$ \\
 ($\gamma,n$) & Sumiyoshi \etal \protect{\cite{Sumi}} & 1.735 & $ 225 \pm 12 $ & 0.568  & $ 0.104 \pm 0.002 $ \\
 ($\gamma,n$) & Present & $ 1.731 \pm 0.002 $ & $ 213 \pm 6 $ & $ 0.738 \pm 0.002 $ & $ 0.136 \pm 0.002$ \\
\hline\hline
\end{tabular}
\end{center}
\label{tab:onehalfparams}
\end{table*}
Notice that all but one of the evaluated virtual photon ($e,e'$) data produced reduced transition strengths and \gray~partial widths that are about half of the value of their real photon \gn~counterparts.  The anomalous parameters within the ($e,e'$) subset of the data~\cite{Barker2000} result from a reanalysis of the cross-section data of Ref.~\cite{Kuechler87}.  Indeed, an inspection of the cross-section data shown in (b) of Fig. \ref{fig:WD_Thresh} from all the works mentioned in Table~\ref{tab:onehalfparams} reveals that the reported cross sections are remarkably similar:  the maximum difference between cross sections is less than a factor of two; cross sections obtained separately from real and virtual photon experiments only vary by 20\% to 40\%. 
It thus appears that different methods of data interpretation, rather than cross-section determinations, give rise to the difference in reported resonance parameters. 
The main difference between the analyses of Refs. \cite{Kuechler87, Burda} and the present analysis involves the use by the former of Siegert's theorem for extracting the B(E1)$\downarrow$. 


\section{\label{sec:ratecalc}\rthrbrHEAD~rate calculation}




\subsection{\label{sec:Reverserxn}Reverse reaction cross section}
The \twobr~ cross section is transformed into the \rtwobr~ cross-section using the reciprocity theorem. Defining $\sigma_1$ to be the cross section for \reverse, and $\sigma_2$ to be the cross section for \forward, the reciprocity theorem gives
\begin{equation}
	\sigma_1 = \frac{2(2j_{^{9} \text{Be} } +1)}{(2j_{^{8} \text{Be} }+1) (2j_{n}+1)}\frac{k_{\gamma}^{2}}{k_{n}^{2}}\sigma_2,
\end{equation}
where $k_n^2 = 2\mu E_n \hbar^{-2}$; 
$k_\gamma^2$ is defined in Eq. \ref{eq:ksubgammasquared}; and the ground state spins for $^{9}$Be, $^{8}$Be, and a neutron are 3/2, 0, and 1/2, respectively.

For several years prior to 1999, the \rthrbr~rate used in reaction network codes was adopted from Ref.~\cite{cf88}, which considered resonant-only decays of \forward.  In other words, when considering the \reverse~ direction for the reaction, the width of the ground state of \bee~was disregarded. The rate published by Ref.~\cite{NACRE} (known as NACRE) included the off-resonant contributions to the \rthrbr~rate. Other rates \cite{Sumi, Burda} have since followed the formalism developed by NACRE. 

\subsection{\label{sec:aanRateCalc}Rate calculation}
The derivation of astrophysical reaction rates has been described in detail in Ref.~\cite{Iliadisbook}. 
The rate per particle pair \sv~is given by
\begin{equation}
	\svmm = \left(\frac{8}{\pi\mu}\right)^{\frac{1}{2}}\left(\frac{1}{k T}\right)^{\frac{3}{2}}\dintfree^{\infty}_{0}{\sigma(E)\exp\left[\frac{-E}{k T}\right]EdE},
\label{eq:normalrate}
\end{equation}
where $\mu$ is the particle pair reduced mass, $k$~is Boltzmann's constant, and $T$ is the temperature.
Equation~\ref{eq:normalrate} is the appropriate form for a two-body reaction. However, here the rate has to be computed for two sequential reactions. 
The form for calculating the rate of formation of \ben~involves constructing a double integral, taking into account the rate of formation of \bee~from the $\alpha + \alpha$ scattering cross sections.  The formalism adopted in the present work was developed in Ref.~\cite{Nomoto85} for calculating the on- and off-resonant formation of \sup{12}C via the triple-$\alpha$ reaction, and was first modified in Ref.~\cite{NACRE} to calculate the rate of formation of \ben~for the NACRE compilation.  

Two \ap s interact with center-of-mass (CM) energy $E$ to form \bee. Subsequently, the \bee~nucleus interacts with a neutron with new CM energy $E'$ relative to $E$ (see Fig.~\ref{Fig:levelscheme}).  The rate equation has the form

\begin{eqnarray}
N^{2}_{A}\left\langle \sigma v \right\rangle^{\alpha\alpha n} = N_{A}\left(\frac{8\pi\hbar}{\mu^{2}_{\alpha\alpha}}\right)\left(\frac{\mu_{\alpha\alpha}}{2\pi kT}\right)^{3/2}\times\nonumber\\
\dintfree^{\infty}_{0}\frac{\sigma_{\alpha\alpha}(E)}{\Gamma_{\alpha}(^{8} \text{Be} ,E)}\exp(-E/kT) N_{A}\left\langle \sigma v \right\rangle^{n^{8} \text{Be} }EdE,
\label{eqn:all}
\end{eqnarray}
with
\begin{eqnarray}
N_{A}\left\langle \sigma v \right\rangle^{n^{8} \text{Be} } = N_{A}\left( \frac{8\pi\hbar}{\mu^{2}_{n^{8} \text{Be} }}\right)\left(\frac{\mu_{n^{8} \text{Be} }}{2\pi kT}\right)^{3/2}\nonumber\\
\times\dintfree^{\infty}_{0}{\sigma_{n^{8} \text{Be} }(E';E)\exp(-E'/kT)E'dE'}.
\label{eqn:npart}
\end{eqnarray}

\begin{figure}[]
  \centering
  \includegraphics[width=.45\textwidth]{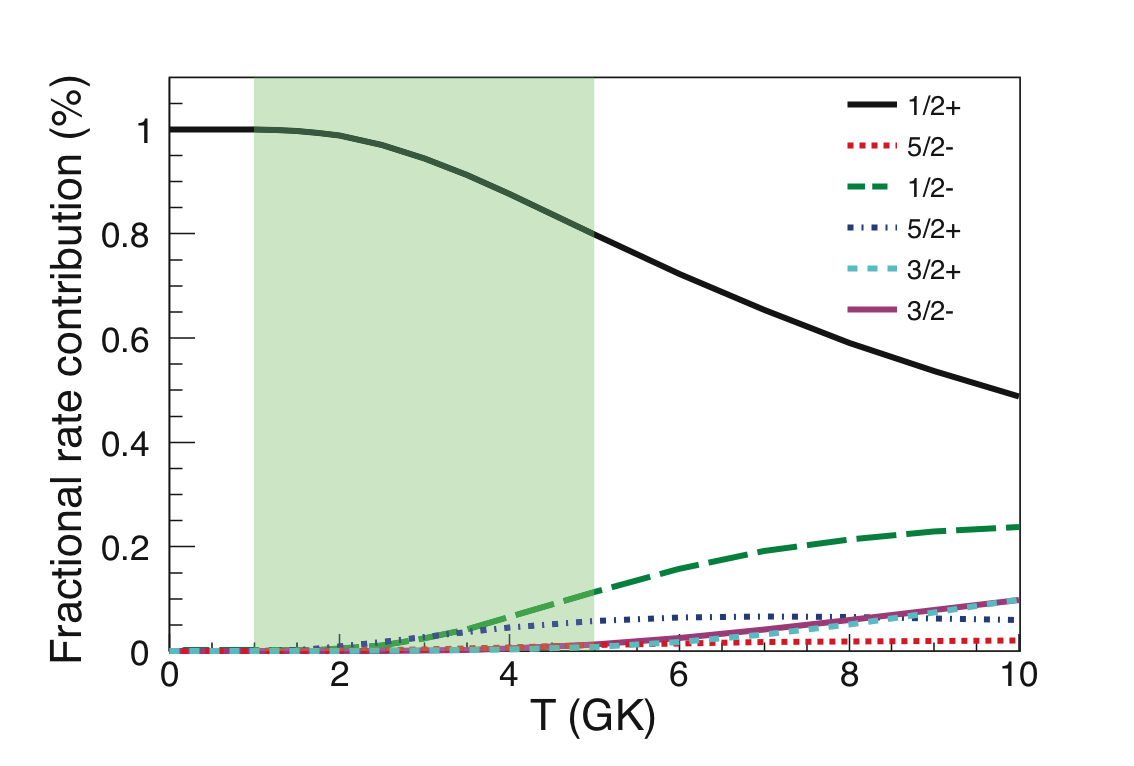}
  \caption{\label{Fig:indy_cont_comp}(Color Online) Relative contributions of each resonance to the total rate as a function of temperature.  The temperature range relevant for the r-process,  1 GK $< T < 5$ GK, is shaded (green).}
\end{figure}

Equation~\ref{eqn:all} is evaluated numerically using the parameters from Table \ref{tab:results} and the \apa~scattering cross-sections from Ref. \cite{Wus92} over the temperature range 0.001 GK $ \leq T \leq 10$ GK. 
Table~\ref{tab:ratesvstemps} lists low, recommended, and high values for the presently determined rates versus temperature. The low and high rates are computed by considering the systematic uncertainty in the deconvoluted \twobr~cross section and the uncertainty in the fitting parameters (see Table~\ref{tab:results}).  At all temperatures, the \rthrbr~rate originates primarily from the 1/2$^{+}$ state in \ben~(see Fig.~\ref{Fig:indy_cont_comp}). Thus, the rate uncertainty is dominated by the cross-section uncertainty for that state. Uncertainty in the rate increases at higher temperatures, where higher states in \ben~begin to contribute noticeably to the \rthrbr~reaction rate.

\begin{longtable*}[c]{ c|ccc|c|ccc }
\caption[]{Low, recommended, and high rates for the \rthrbr~reaction versus $\text{T}_{9}$ ($\equiv 1$ GK ) computed from the present parameters.}
\label{tab:ratesvstemps}\\
\hline\hline
~       &~   & N$_{A}$\aan &~   &~        &~   & N$_{A}$\aan & \\
T$_{9}$ & Low& Recommended    &High& T$_{9}$ & Low& Recommended    &High \\
\hline
\endfirsthead
\caption[]{(Continued)}\\
~       &~   & N$_{A}$\aan &~   &~        &~   & N$_{A}$\aan & \\
T$_{9}$ & Low& Recommended    &High& T$_{9}$ & Low& Recommended    &High \\
\hline
\endhead
0.001&1.15E-59&1.20E-59&1.26E-59&0.14&4.39E-08&4.62E-08&4.84E-08\\
0.002&9.64E-48&1.01E-47&1.06E-47&0.15&6.49E-08&6.82E-08&7.14E-08\\
0.003&6.01E-42&6.32E-42&6.62E-42&0.16&9.05E-08&9.51E-08&9.96E-08\\
0.004&2.76E-38&2.90E-38&3.04E-38&0.18&1.54E-07&1.62E-07&1.70E-07\\
0.005&1.12E-35&1.18E-35&1.24E-35&0.2&2.32E-07&2.43E-07&2.55E-07\\
0.006&1.11E-33&1.17E-33&1.23E-33&0.25&4.49E-07&4.72E-07&4.95E-07\\
0.007&4.40E-32&4.62E-32&4.84E-32&0.3&6.51E-07&6.84E-07&7.17E-07\\
0.008&9.21E-31&9.67E-31&1.01E-30&0.35&8.07E-07&8.48E-07&8.89E-07\\
0.009&1.21E-29&1.27E-29&1.33E-29&0.4&9.10E-07&9.57E-07&1.00E-06\\
0.01&1.12E-28&1.18E-28&1.24E-28&0.45&9.70E-07&1.02E-06&1.07E-06\\
0.011&7.90E-28&8.31E-28&8.71E-28&0.5&9.94E-07&1.04E-06&1.10E-06\\
0.012&4.47E-27&4.70E-27&4.93E-27&0.6&9.76E-07&1.03E-06&1.07E-06\\
0.013&2.12E-26&2.22E-26&2.33E-26&0.7&9.10E-07&9.56E-07&1.00E-06\\
0.014&8.64E-26&9.08E-26&9.52E-26&0.8&8.29E-07&8.71E-07&9.13E-07\\
0.015&3.12E-25&3.28E-25&3.43E-25&0.9&7.45E-07&7.83E-07&8.21E-07\\
0.016&1.01E-24&1.06E-24&1.12E-24&1&6.66E-07&7.00E-07&7.34E-07\\
0.018&8.24E-24&8.65E-24&9.07E-24&1.25&5.02E-07&5.28E-07&5.53E-07\\
0.02&5.09E-23&5.35E-23&5.61E-23&1.5&3.83E-07&4.02E-07&4.22E-07\\
0.025&2.51E-21&2.64E-21&2.77E-21&1.75&2.98E-07&3.13E-07&3.28E-07\\
0.03&4.31E-19&4.53E-19&4.75E-19&2&2.37E-07&2.49E-07&2.61E-07\\
0.04&1.81E-15&1.90E-15&2.00E-15&2.5&1.58E-07&1.66E-07&1.74E-07\\
0.05&2.63E-13&2.76E-13&2.90E-13&3&1.12E-07&1.18E-07&1.24E-07\\
0.06&6.88E-12&7.23E-12&7.58E-12&3.5&8.43E-08&8.88E-08&9.33E-08\\
0.07&6.81E-11&7.15E-11&7.50E-11&4&6.59E-08&6.95E-08&7.32E-08\\
0.08&3.68E-10&3.87E-10&4.06E-10&5&4.41E-08&4.67E-08&4.94E-08\\
0.09&1.34E-09&1.40E-09&1.47E-09&6&3.22E-08&3.43E-08&3.65E-08\\
0.1&3.67E-09&3.86E-09&4.05E-09&7&2.49E-08&2.68E-08&2.86E-08\\
0.11&8.27E-09&8.69E-09&9.11E-09&8&2.02E-08&2.18E-08&2.35E-08\\
0.12&1.60E-08&1.68E-08&1.77E-08&9&1.69E-08&1.84E-08&1.99E-08\\
0.13&2.77E-08&2.92E-08&3.06E-08&10&1.46E-08&1.59E-08&1.74E-08\\ 
\hline\hline
\end{longtable*}

\begin{figure}[] 
  \centering
  \includegraphics[width=.49\textwidth]{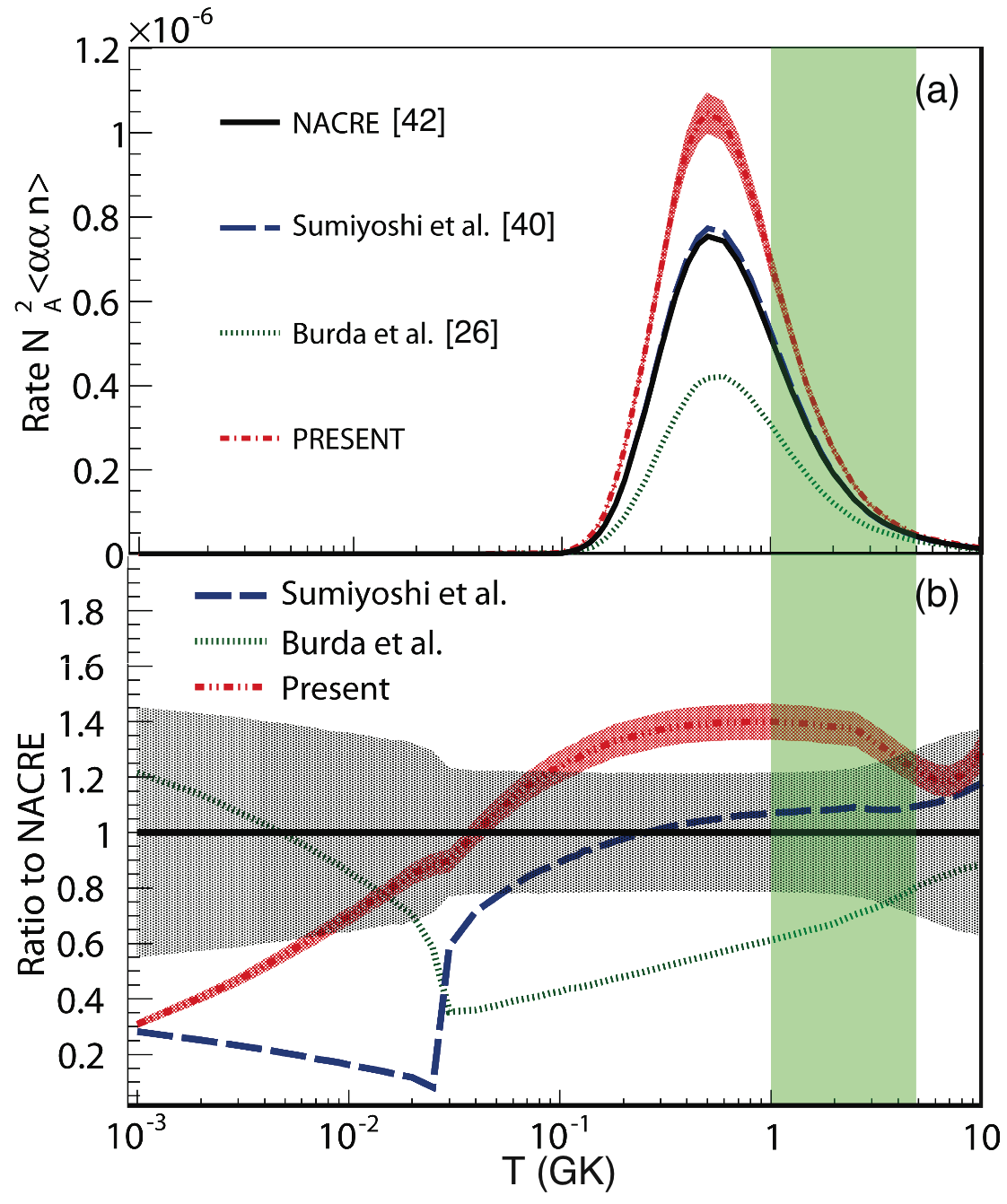}
  \caption{\label{Fig:Twopannel}(Color Online)(a) The reaction rate as a function of temperature. The present rate is 40\% larger than the rates of Refs. \protect{\cite{Sumi}} and \protect{\cite{NACRE}} at the peak near $T = 0.5 ~ \text{GK}$. The temperature range relevant for the r-process,  1 GK $< T < 5$ GK, is shaded (green).
(b) The ratio of the rates to the NACRE rate.  Bands shown indicate $\pm 1 \sigma$ uncertainties for NACRE and the present rate.}
\end{figure}

Figure~\ref{Fig:Twopannel} displays the comparison of four existing rates both by their absolute magnitudes and by normalization to the NACRE rate ~\cite{NACRE}.  
The present rate is 40\% larger than the NACRE result for the energy range 1 GK $ < T < 5$ GK, most important for r-process nucleosynthesis.  The largest difference between the various rates exists in the off-resonant region: there the present rate is smaller than the NACRE rate by a factor of 4 because the energy dependence of all resonant cross-section contributions near the two-body threshold has been included.



\subsection{Rate comparisons}
In principle, the precision of the measured \twobr~reaction cross section should extend to the deduced astrophysical \rthrbr~rate.  The quantities used for calculating the \apa~$\rightarrow$ \bee~rate derive from well known \apa~scattering data~\cite{Wus92}, while accurate penetration factors are obtained from computed Coulomb wave functions.
Recalling Fig. \ref{Fig:Twopannel}, the present rate, when compared to the NACRE rate, is a factor of 3 lower at the lowest calculated temperatures, while it is 20\% to 40\% larger at astrophysical temperatures of interest for the r-process (1 GK $ < T < 5$ GK).  The change of the low-temperature rate is a direct result of including a realistic energy dependence for all neutron partial widths, a procedure which was not employed in Ref.~\cite{NACRE}.  The correct analytic form for \textit{s}-, \textit{p}-, and \textit{d}-wave neutron penetration factors lowers all resonant contributions near threshold as shown in Fig. \ref{Fig:thresh_behav}. 
 At low temperatures, the energy-independent treatment leads to an artificially inflated rate along with the non-physical dominant contributions to this rate by the \textit{p}-wave $1/2^-$ and \textit{d}-wave $5/2^+$ states.
 
 Recently, Garrido \emph{et al.} \cite{garrido} have used three-body theoretical techniques to determine the direct \aantoben~contribution to both the total cross section and reaction rate at energies below the two-body threshold. However, their calculated direct, three-body cross section exceeds the 93 nb experimental upper limit \cite{threebody} at energies between the three- and two-body thresholds. The results presented in the present paper consider only sequential reactions and do not address the possibility of contributions by three-body processes at the lowest temperatures. 

The present \rthrbr~rate  for 1 GK $ \leq T \leq 5$ GK is consistently 20\% to 40\% larger than the rates of Refs.~\cite{NACRE, Sumi}. Agreement with the NACRE rate marginally improves as the temperature approaches 10 GK.  Figure \ref{Fig:indy_cont_comp} shows that, for the present evaluation, the 1/2$^{+}$ state is indeed the primary contributor to the rate at temperatures below 10 GK.  Contributions from other states start to become noticeable for $T \geq 5$ GK.  This implies that the choice of branching ratios is not important for an accurate \rthrbr~rate determination.  

\section{\label{sec:discuss}Conclusions}
The improved accuracy of the measurement of the \twobr~reaction cross section reported in this work was made possible: a) by use of a highly-efficient neutron detector with two concentric, circular arrays of $^3$He tubes which provided information about the energy distribution of the detected neutrons; b) by calibration of the neutron detector's efficiency using interspersed measurement of the well-known $^2$H($\gamma$,n) cross section; and c) by measuring at each energy both the flux and energy distribution of the incident $\gamma$-ray beam used. Knowledge of the incident beam's energy distribution allowed deconvolution and determination of the cross section near the neutron threshold at 1665 keV.
These new measurements have been used to calculate the astrophysical \rthrbr~reaction rate. 

Taking into account energy dependence of all neutron and \gray~partial widths near threshold gives rise to smaller rates than previously calculated for $T \leq 0.025$ GK. For $T \geq 2$ GK, contributions to the rate from higher-lying resonances become noticeable.  Our cross sections and the resulting astrophysical reaction rates in the temperature range 1 GK $ \leq T \leq 5$ GK are 20\% to 40\% larger than previously reported. The present rate is computed using a cross section known to $\pm 10\%$ at the 95\% confidence level.  This new rate should be employed in reaction network codes, especially those used to investigate r-process nucleosynthesis sites. 

\begin{acknowledgments}
This work was supported in part by USDOE Office of Nuclear Physics Grants DE-FG02-97ER41041 and DE-FG02-97ER41033.  
We wish to acknowledge the staff of the UNC Chapel Hill and Duke instrument shops; experimental collaborators G. Rusev, S. Stave, M.W. Ahmed, and Y. Wu;  and the \higs~accelerator staff for their help in completing these measurements.
\end{acknowledgments}

\bibliography{be9gn_for_prc}

\end{document}